\documentclass[twocolumn,pra]{revtex4}

\usepackage{amsmath}
\usepackage{amstext}
\usepackage{latexsym}
\usepackage{psfig}
\usepackage{graphicx}
\usepackage{amsfonts}

\begin{document}
%\draft

%\wideabs{

\title{
Generation of macroscopic superposition states
 with small nonlinearity}

\author{$^1$H. Jeong, $^2$M.S. Kim, $^1$T.C. Ralph, and $^3$B.S. Ham}

\address{$^1$Centre for Quantum Computer Technology, Department of Physics, University of Queensland, St Lucia, Qld 4072,
   Australia \\
$^2$School of Mathematics and Physics, Queen's University,
Belfast BT7 1NN, United Kingdom \\
$^3$Graduate School of Information \& Communication,
Inha University, Incheon, South Korea
}

\date{\today}

\begin{abstract}
We suggest a scheme to generate a macroscopic superposition state
(Schr\"odinger cat state) of a free-propagating optical field
using a beam splitter, homodyne measurement and a very small
Kerr nonlinear effect. Our scheme makes it possible to considerably
reduce the required nonlinear effect to generate an optical cat state
using simple and efficient optical elements.
\end{abstract}
%\pacs{PACS number(s);} %}% 03.67.-a, 89.70.+c}  }
\maketitle

{\it Introduction}-- 
In the well-known cat paradox, 
Schr\"odinger tried to demonstrate a possibility of generating
a quantum superposition of a macroscopic system \cite{Sch}.
A superposition of two coherent states with a $\pi$ phase
difference and a large amplitude is considered a
realization of such a macroscopic superposition
and sometimes called ``Schr\"odinger cat state". Recently,
it has been found that the cat state of a propagating optical
field is useful not only for the study of fundamental quantum
physics but also for various applications to quantum information
processing \cite{Dur,Enk,JKL01,Wang,JK,Ralph,JKpuri,MunroForce}.
Once the optical cat state is generated, quantum teleportation
\cite{Enk,JKL01,Wang}, quantum nonlocality test \cite{Derek},
generation and purification \cite{JKL01,JKpuri} of entangled
coherent states, quantum metrology  \cite{MunroForce}, and
quantum computation \cite{JK,Ralph,Ralph2} will become closer to
experimental realization using current technology.

It has been theoretically known that the cat state can be
generated from a coherent state by a nonlinear interaction in a
Kerr medium \cite{Yurke}. However, the Kerr nonlinearity of
currently available media is too small to generate the 
cat state.  It was pointed out that one needs
an optical fiber of about 1,500km for an optical frequency of
$\omega\approx5\times10^{14}$rad/sec to generate a coherent
superposition state with currently available Kerr nonlinearity
\cite{Gerry,SaMi92}. Even though it is possible in principle to
make such a long nonlinear optical fiber, the 
effects of decoherence and phase fluctuations
during the propagation will become too large. 

Some alternative methods have been studied to generate a
superposition of macroscopically distinguishable states using
conditional measurements \cite{Song,Dakna}. One drawback of these
schemes is that a highly efficient photon counting measurement is
required to obtain a coherent superposition state, which is
difficult using current technology.  Cavity quantum
electrodynamics has been studied to enhance nonlinearity
\cite{Tu}. Even though there have been experimental
demonstrations of generating cat states in a cavity and in a trap
\cite{MB,Mon}, all the suggested schemes for quantum information
processing with coherent states
\cite{Enk,JKL01,Wang,JK,Ralph,JKpuri} require {\it free
propagating} optical cat states.

Electromagnetically induced transparency (EIT) has been studied
as a method to obtain a giant Kerr nonlinearity \cite{giant}.
There has been an inspiring suggestion to generate cat states
with it \cite{Mauro} but this developing technology of EIT has
not been exactly at hand yet to generate a state in a quantum
regime.  Recently, Lund {\it et al.} proposed a simpler optical
scheme to generate a propagating cat state of $|\alpha|\approx2$
\cite{Lund}, which does not require Kerr-type nonlinearity
nor photon counting measurements.

In this paper, we study a probabilistic scheme to generate cat
states with a small Kerr effect. We are particularly interested in
generating a cat state of $|\alpha|\geq10$, {\it i.e.}, the average
photon number over 100. Cat states with large amplitudes are preferred
for quantum information processing. For example, higher precision is
obtained for quantum metrology when large cat states are supplied
\cite{MunroForce}. Our scheme significantly reduces the required
nonlinear effect to generate cat states using a beam splitter and
homodyne measurement which are basic and efficient tools in
quantum optics laboratories.

{\it Generating a cat state with Kerr nonlinearity and its
limitation}-- A cat state is defined as
\begin{equation}
\label{CaT}
|cat_{\alpha,\varphi}\rangle={\cal N}(\alpha,\varphi)(|\alpha\rangle
+e^{i\varphi}|-\alpha\rangle),
\end{equation}
where ${\cal N}(\alpha,\varphi)$ is a normalization factor,
$|\alpha\rangle$ is a coherent state of amplitude $\alpha$,
and $\varphi$ is a real local phase factor. Note that the relative
phase $\varphi$ can be approximately controlled by the
displacement operation for a given cat state with $\alpha\gg1$
\cite{JK,Cochrane}. The Hamiltonian of a single-mode Kerr
nonlinear medium is 
${\cal H}_{NL}=\omega a^\dag a+\lambda(a^\dag a)^2$, 
where $a$ and $a^\dag$ are annihilation and creation operators,
$\omega$ is the energy level splitting for the
harmonic-oscillator part of the Hamiltonian and $\lambda$ is the
strength of the Kerr nonlinearity \cite{Yurke}. Under the influence of the
nonlinear interaction the initial coherent state with the
coherent amplitude $\alpha$ evolves to the following state at
time $\tau$:
\begin{equation}
\label{c1} |\psi(\tau)\rangle=e^{-|\alpha^2|/2} \sum_{n=0}^\infty
\frac{\alpha^n e^{-i\phi_n}}{\sqrt{n!}}|n\rangle,
\end{equation}
where $\phi_n=\lambda\tau n^2$.
When the interaction time $\lambda\tau$ in the medium is $\pi/
N$ with a positive integer $N$, the initial coherent state $|\alpha\rangle$ evolves to
\cite{LKLB}
\begin{equation}
\label{c2}
|\psi_N\rangle=\sum_{n=1}^N C_{n,N}|-\alpha
 e^{2in\pi/N}\rangle,
\end{equation}
where $C_{n,N}=e^{i\zeta_n}/\sqrt{N}$.  Comparing Eqs.~(\ref{c1})
and (\ref{c2}) for an arbitrary $N$, we find an equation for the
arguments $\zeta_n$'s of the coefficients of the coherent
components, {\em i.e.},
\begin{equation}
\label{NC}
\frac{1}{\sqrt{N}}\sum_{n=1}^N e^{i\zeta_n} (-e^{2in\pi/N})^k = \exp(-i\pi k^2/N).
\end{equation}
By solving the $N$ coupled equations given by Eq.~(\ref{NC}), the
values $\zeta_n$'s  are obtained as \cite{SR99}
\begin{equation}
C_{n,N}=\frac{e^{i\zeta_n}}{\sqrt{N}}=\frac{1}{N}\sum_{k=0}^{N-1}(-1)^k
\exp[-\frac{i\pi k}{N}(2n-k)].
\end{equation}
The process shown above can produce a
large amount of entanglement in a short time
\cite{enk-inf}.
 The length $L$ of the nonlinear cell
corresponding to $\tau$ is $L=v\pi/2\lambda N$, where $v$ is the
velocity of light.  For $N=2$, we obtain a desired cat state of
the form (\ref{CaT}) with $\varphi=\pi/2$ \cite{Yurke}.
We pointed out that the nonlinear coupling $\lambda$ is typically
very small such that $N=2$ cannot be obtained in a length limit where
the decoherence effect can be neglected.

\begin{figure}
\centerline{\scalebox{0.58}{\includegraphics{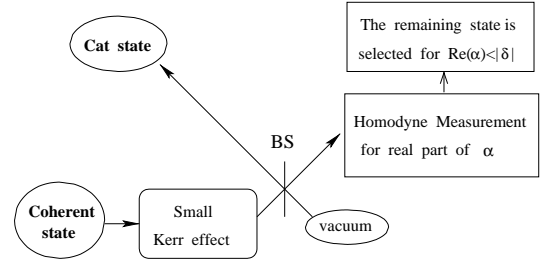}}}
 \caption{Schematic of generating a cat state using
small Kerr nonlinearity, a beam splitter and a homodyne
detection. } \label{fig:scheme}
\end{figure}

{\it Generating a cat state with smaller nonlinearity}-- If
$\lambda$ is not as large as required to generate the cat state,
the state (\ref{c2}) with $N>2$ may be obtained
by choosing an appropriate interaction time.
  From the state (\ref{c2}),
it is required to remove all the other coherent component states
except two coherent states of a $\pi$ phase difference.  First,
we assume that the state (\ref{c2}) is incident on a 50-50 beam
splitter with the vacuum onto the other input of the beam
splitter, as shown in Fig.~\ref{fig:scheme}.
 The initial coherent
amplitude $\alpha_i$ is taken to be real for simplicity. The
state (\ref{c2}) with initial amplitude $\alpha_i$ after passing
through the beam splitter becomes
\begin{equation}
\label{cbv}
|\psi_N\rangle=\sum_{n=1}^N C_{n,N}|-\alpha_i e^{2in\pi/N}/\sqrt{2}\rangle
|-\alpha_i e^{2in\pi/N}/\sqrt{2}\rangle,
\end{equation}
where all $|C_{n,N}|$'s have the same value. The real part
of the coherent amplitude in the state (\ref{cbv}) is then
measured by homodyne detection
in order to produce the cat state in the other path.
 By the measurement result, the
state is reduced to
\begin{eqnarray}
\label{cm} |\psi^{(1)}_N\rangle= \sum_{n=1}^N
C^{(1)}_{n,N}(\alpha_i) |-\alpha_i e^{2in\pi/N}/\sqrt{2}\rangle,
\end{eqnarray}
where $C^{(1)}_{n,N}(\alpha_i) = {\cal N}_\psi \sum_{n=1}^N
C_{n,N } \langle X|-\alpha_i e^{2in\pi/N}/\sqrt{2}\rangle$ with
${\cal N}_\psi$ the normalization factor and $|X\rangle$ the
eigenstate of ${\hat X}=(a+a^\dagger)/\sqrt{2}$.
 After the
homodyne measurement, the state is selected when the measurement
result is in certain values.  If coefficients
$|C^{(1)}_{N/2,N}(\alpha_i)|$ and $|C^{(1)}_{N,N}(\alpha_i)|$ in
Eq.  (\ref{cm}) have the same nonzero value and all the other
$|C^{(1)}_{n,N}(\alpha_i)|$'s are zero, then the state becomes a
desired cat state. Suppose $N=4k$ where $k$ is a positive integer number.
If $X=0$ is measured in this case, the
coefficients $|C^{(1)}_{n,N}(\alpha_i)|$'s will be the largest
when $n=N/4$ and $n=3N/4$, and become smaller as $n$ is far from these
two points. The coefficients can be close to zero for all the
other $n$'s for an appropriately large $\alpha_i$ so that the
resulting state may become a cat state of high fidelity.

The fidelity between the state (\ref{cm}) obtained by our process
and a `perfect' cat state of the form (\ref{CaT}) with appropriate amplitude is
\begin{eqnarray}
&&f(\alpha_i,N,X)=\max_\varphi\Big[|\langle cat_{\alpha_i/\sqrt{2},\varphi}
|\psi^{(1)}_N\rangle|^2\Big]\nonumber\\
&&=\max_\varphi\Big[{\cal N}(\alpha_i,\varphi)^2{\cal N}_\psi^2\Big|
\sum_{n=1}^N C_{n,N}^{(1)}
\exp[-\frac{\alpha_i^2}{2}(1+e^{2 i n \pi/N})] \nonumber\\
&&~~~~~~~+   
e^{i\varphi}\sum_{n=1}^N C_{n,N}^{(1)}
\exp[-\frac{\alpha_i^2}{2}(1-e^{2 i n \pi/N})]
\Big|^2\Big].
\end{eqnarray}
The success probability to get a cat state is 
\begin{eqnarray}
&&{\cal P}(\alpha_i,N,\delta)=\int_\delta dX{\rm  Tr}[\rho_1|X\rangle\langle X|] \nonumber\\
&&=\int_\delta dP\sum^N_{nm} \langle-\alpha_i
e^{2in\pi/N}/\sqrt{2}
|X\rangle\langle X|-\alpha_i e^{2im\pi/N}/\sqrt{2}\rangle \nonumber \\
&&~~~~~~~~~~\times\exp[-\frac{\alpha_i^2}{2}(1-e^{2 i (m-n)
\pi/N})] \label{eq--prob}
\end{eqnarray}
where $\rho_1={\rm Tr}_2[|\psi_N\rangle_{12}~_{12}\langle\psi_N|]$ and
$\delta$ is the range in which the high fidelity is obtained.  Note
that the initial coherent amplitude $\alpha_i$ needs to be larger as $N$
increases for better fidelity.

\begin{figure}
\centerline{\scalebox{0.58}{\includegraphics{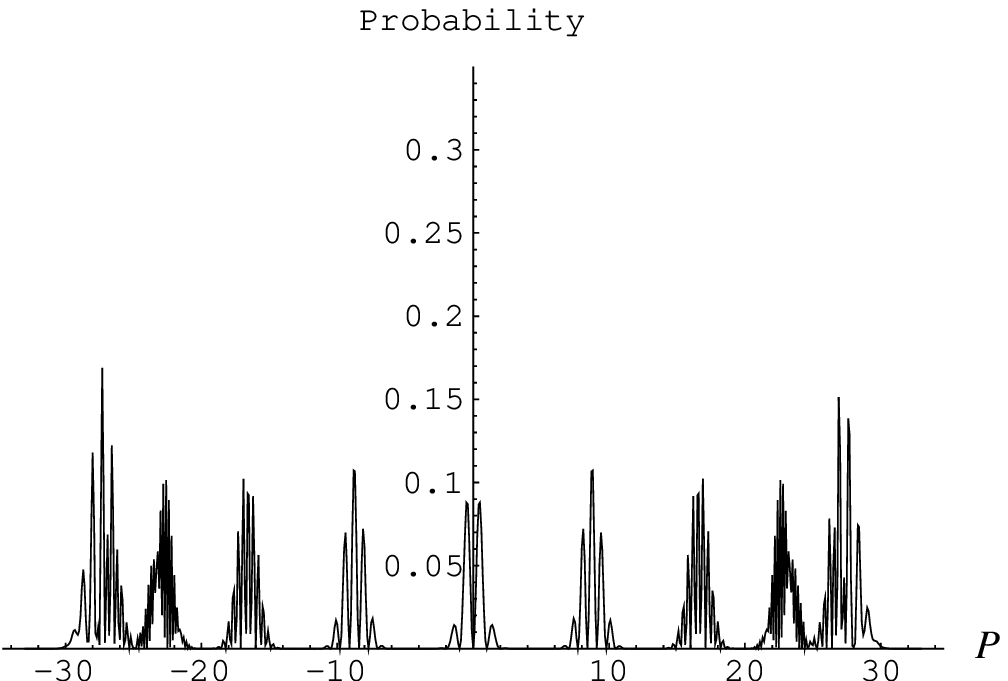}}}
\centerline{\scalebox{0.58}{\includegraphics{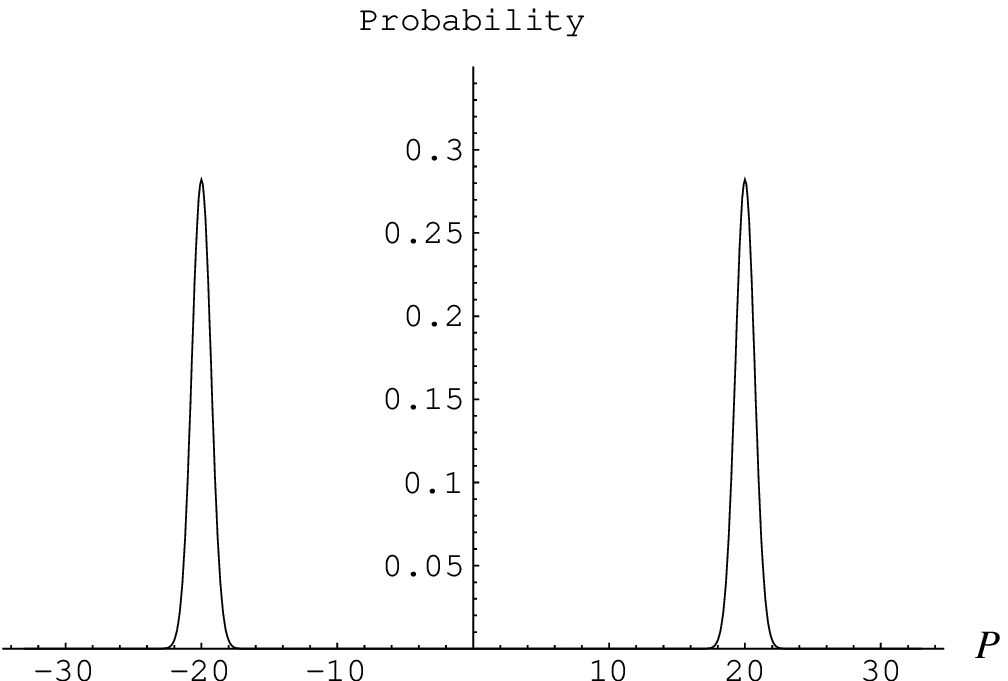}}}
 \caption{(a) The probability distribution of 
 the quadrature variable $P$ for 
 the state
after passing the small nonlinear medium but {\it before}
the beam splitter in Fig.~1.
The interaction
time $\lambda\tau=\pi/20$ and the initial amplitude $\alpha_i=20$.
(b) The probability distribution for the state
{\it after} the beam splitter and homodyne measurement 
of result $X=0$.
 It is obvious from the
figures that a cat state with well separated peaks is obtained
after the process. } \label{fig:prob}
\end{figure}

\begin{figure}
\centerline{\scalebox{0.58}{\includegraphics{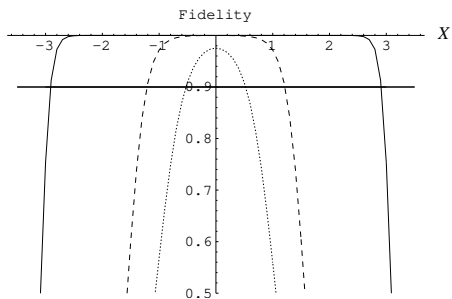}}}
\caption{Fidelity $F$ of the generated cat state against the measurement outcome $X$
where $N=20$ (solid line), $N=40$ (dashed line), $N=60$ (dotted line). 
The initial amplitude is $\alpha_i=20$.}
\label{fig:fidelity}
\end{figure}

\begin{figure}
\centerline{\scalebox{0.58}{\includegraphics{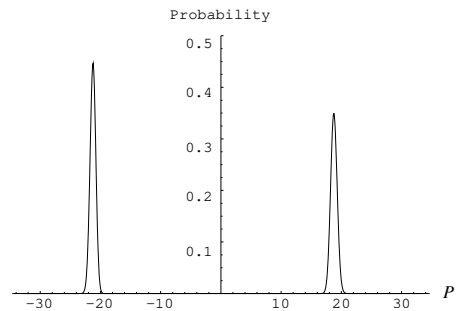}}}
\caption{The probability distribution of $P$ for the state
conditioned on the homodyne measurement 
of result $X=0$ after the beam splitter when the interaction time
is $\lambda=\pi/200$ and the initial amplitude is $\alpha_i=20$}
\label{fig:p200}
\end{figure}

We first examine an example of $\alpha_i=20$ and
$\lambda\tau=\pi/20$, {\it i.e.}, the interaction time (or the
nonlinear strength) is an order of magnitude shorter (weaker) than the required
value. After passing through the nonlinear medium, the fidelity
between the generated state and an ideal cat state is
$F\approx0.1$. The probability distribution of $P$, which is the
conjugate variable of $X$, is shown in Fig.~\ref{fig:prob}(a).
After beam splitting and the homodyne measurement are applied,
the state is drastically reduced to a cat state with amplitude
$|\alpha|=\alpha_i/\sqrt{2}\approx14.1$. The maximum fidelity of
this cat state is when the measurement result is $X=0$ for
$\varphi=\pi$. Fig.~\ref{fig:prob}(b) shows two well separated
peaks of the cat state produced for the case of $X=0$.  A high
fidelity 
 is obtained for a certain range $\delta$ of the
measurement outcome as shown in Fig~\ref{fig:fidelity}. The total
success probability can be calculated by integrating
Eq.~(\ref{eq--prob}) over $\delta$.  The success probability for
$F>0.99999$  is numerically obtained as $\approx10\%$, which means
that only 10 trials are required on average to obtain a cat
state of $F>0.99999$.
The success probability for $F>0.99$
is about 4\% if $\lambda\tau=\pi/40$.
If $\lambda\tau=\pi/60$, the maximum fidelity is 0.975 and
the success probability for $F>0.9$ is ~2\%.
Our scheme generates well separated peaks even for $\lambda\tau=\pi/200$
as shown in Fig.~\ref{fig:p200}.
Even though the resulting state in this case is somewhat different from the 
ideal cat state in Eq.~(\ref{CaT}),
this shows that one may observe a conspicuous signature of Schr\"odinger
cat state (macroscopic superposition) even with a 1/100 times weaker nonliearity
compared with the currently required level.

{\it Error sources}--
In our proposal, a coherent state can be well approximated by
a laser field. Beam splitting of a propagating field
with a vacuum can also be efficiently performed 
using current technology \cite{Pitt}.
However, decoherence and random phase fluctuations
in a nonlinear medium would be main sources
of errors in our scheme. We have neglected
these effects because the initial state would pass
through a relatively short length of medium. 
We now roughly assess the effects of decoherence and random
phase fluctuations as follows.

Besides the gradual reduction of amplitude by the loss of the average energy,
photon losses will cause the loss of the phase information and make
the final state mixed. An analysis of Eqs.~(3) to (6) shows that 
photon losses only at the later stage in the nonlinear medium 
will significantly affect the phase of the final state.
We assume that photon losses only at the final stage
change the phase. In this case, if an odd number of photons
are lost the phase of the final state is filpped by $\pi$ 
while it does not change if an even number of photons are lost.
The final state is then represented by
\begin{equation}
\rho_{env}^{(1)}=(1-{\cal P}_f)|\Psi_{N}^{(1)} \rangle\langle\Psi_{N}^{(1)}|
+{\cal P}_f|{\Phi}_{N}^{(1)}\rangle\langle\Phi_{N}^{(1)}|,
\end{equation}
where ${\cal P}_f=\sum_{n=0}^\infty p_{2n+1}$, $p_n$ is the probability of losing $n$ photons,
$|\Psi_{N}^{(1)} \rangle=|{\psi_{N}^{(1)}}_{\alpha_i e^{-\frac{\gamma\tau}{2}},~\varphi}\rangle$,
$|\Phi_{N}^{(1)} \rangle=|{\psi_{N}^{(1)}}_{\alpha_i e^{-\frac{\gamma\tau}{2}},~\varphi+\pi}\rangle$,
and $\gamma$ is the energy decay rate.
The probability $p_n$
is given by a Poisson distribution. 
If the probability of losing photons at the final stage is 
10\%, the maximum fidelity will be $F\approx0.88$.
If the probability increases to  
30\% (60\%), the maximum fidelity will decrease to $F\approx0.71$ ($F\approx0.55$).

\begin{figure}
\centerline{\scalebox{0.58}{\includegraphics{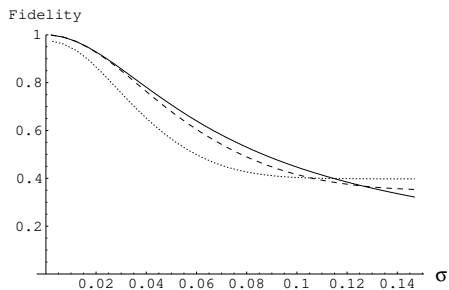}}}
\caption{The average fidelity between the ideal cat state and the 
phase-fluctuated state with standard deviation $\sigma$  
during the process in Fig.~1. The cases for
$N=20$ (solid line), $N=40$ (dashed line) and $N=60$ (dotted line)
 have been plotted with the initial amplitude $\alpha_i=20$. 
The result of the homodyne measurement is considered $X=0$.
}
\label{fig:rp}
\end{figure}

The phase of the state in the medium 
can randomly fluctuate during the process.
For example, if the phase fluctuates by $\Delta\varphi$ during the process,
the final state will be
$|\psi_N^{\Delta\varphi}\rangle\approx\sum_{n=1}^N \
C^{(1),\Delta\varphi}_{n,N}
|-\alpha_i e^{i(2n\pi/N+\Delta\varphi)}/\sqrt{2}\rangle$,
where $C^{(1),\Delta\varphi}_{n,N}=C_{n,N}\langle X|-\alpha_i e^{i(2n\pi/N+\Delta\varphi)}/\sqrt{2}\rangle$.
We suppose the distribution of the phase fluctuation
is Gaussian. The average fidelity between the phase-fluctuated 
state and the ideal cat state
for the measurement result $X=0$ will then be
$F=\int_{-\infty}^{\infty}d \Delta\varphi
G(\Delta\varphi,\sigma)
 \langle cat_{\alpha_i/\sqrt{2},~\varphi_{max}}
 |\psi_N^{\Delta\varphi}\rangle$,
where $\varphi_{max}$ is the fidelity-maximizing
phase for $\Delta\varphi=0$
and $G(\Delta\varphi,\sigma)$ is the Gaussian distribution of $\Delta\varphi$
with standard deviation $\sigma$.
The fidelity against the standard deviation $\sigma$ is
 plotted 
in Fig.~\ref{fig:rp}.
It can be simply shown that the phase fluctuation 
in Fig.~\ref{fig:rp} 
is just the same order of the Gaussian weighted integration of
$|\langle\alpha_i/\sqrt{2}|\alpha_ie^{i\Delta\varphi}/\sqrt{2} \rangle|^2$.
This kind of phase fluctuation problem 
is typical in 
continuous-variable quantum optics experiments
such as a squeezing experiment.
One can significantly reduce this sensitivity by being
less ambitious for making a large cat state, {\it i.e.},
by reducing the amplitude of the initial coherent state.

{\it Remarks}-- We have suggested an optical scheme using a beam
splitter and homodyne detection to generate a cat state with
relatively small nonlinearity. It has been found that the
required nonlinear effect to generate a useful cat state with
$|\alpha|>10$ and $F>0.9$ can be reduced to less than 1/30.
A signature of a Schr\"odinger cat state can be
obtained even with a 1/100 times weaker nonlinearity
compared with the currently required level.

Our scheme is an effort to considerably
reduce the required nonlinear effect to generate a cat state
using a beam splitter and homodyne detection which are
efficient 
tools in quantum optics laboratories. 
Experimental
efforts are being made for optical fibers with loss as low as
0.01db/km where a signal
attenuates by half  in about 300km \cite{foresee,zrf4}.
If one can reduce the required level of nonlinearity by,
{\it e.g.}, 30 times (or 100 times), 
 such a level of nonlinear
effect will be gained in an optical fiber of 50km (or 15km). 
Then there will be a significantly
 improved possibility of producing a cat state using
the nonlinear fiber.
Various nonlinear crystals may be considered instead of optical fibers.
It might be possible to obtain even
lower ratios of losses to nonlinearity  
by using whispering gallery modes of a microsphere
constructed from a nonlinear material \cite{Gal}.

We thank S.J. van Enk and F. K\"onig for their useful comments.
We acknowledge the UK Engineering and Physical Science Research
Council, the Korea Research Foundation (2003-070-C00024), the
Australian Research Council and the University of Queensland
Excellence Foundation.

\end{document}